\documentclass[prd,aps,preprintnumbers, showpacs, nofootinbib,superscriptaddress,notitlepage]{revtex4-1}
\usepackage{mathrsfs}
\usepackage{amsfonts}
\usepackage{amsmath}
\usepackage{array}
\usepackage{verbatim}
\usepackage{epsfig}
\usepackage{graphicx}
\usepackage[dvipsnames]{xcolor}  
\usepackage{graphicx} 
\usepackage[font=small,labelfont=bf]{caption} 
\usepackage[subrefformat=parens]{subcaption}

\newcommand{\beq}{\begin{eqnarray}}
\newcommand{\eeq}{\end{eqnarray}}

\newcommand{\nn}{\nonumber \\}

\begin{document}

\begin{flushright}

\end{flushright}

\title{Spin-flip gluon GTMD $F_{1,2}$ at small-$x$}

\author{Sanskriti Agrawal}
\affiliation{Department of Physics, Aligarh Muslim University, Aligarh - $202001$, India.}

\author{Nahid Vasim}
\affiliation{Department of Physics, Aligarh Muslim University, Aligarh - $202001$, India.}

\author{Raktim Abir}
\affiliation{Department of Physics, Aligarh Muslim University, Aligarh - $202001$, India.}
\email{raktim.ph@amu.ac.in}

\begin{abstract}
Until recently the spin-flip processes in the deep inelastic scatterings are thought to be suppressed in the high energy. 
We found a positive intercept for the spin-flip generalized transverse momentum-dependent parton distribution (GTMDs) ${\rm Re}(F_{1,2})$ as, 
\begin{eqnarray}
{\rm Re}(F_{1,2}) \sim \left(\frac{1}{x}\right)^{{\bar \alpha}_s\left(4\ln2-8/3\right)} \left(\cos 3\phi_{k\Delta} +\cos \phi_{k\Delta}\right). \nonumber
\end{eqnarray}
This is done by analytically solving the integro-differential evolution equation for ${\rm Re}(F_{1,2})$, recently proposed by Hatta and Zhou, in the dilute regime. Interestingly, the surviving solution corresponds to conformal spin $n=2$ and carries an explicit $\cos 3\phi_{k\Delta} + \cos \phi_{k\Delta}$ azimuthal dependence. As the imaginary part of $F_{1,2}$, is related to the spin-dependent odderon or gluon Siver function and scales as ${\rm Im}(F_{1,2}) \sim x^{0}$, the positive intercept for ${\rm Re}(F_{1,2})$, implies that it is expected to dominate over the gluon Siver function in the small-$x$ limit and may directly impact the modeling of unpolarised GTMDs and associated spin-flip processes. 

\end{abstract}

\maketitle

{\color{NavyBlue} {\it Introduction}:~}
%
%
Two key physics goals of the upcoming Electron-Ion Collider are to profile the inner structure of the proton and to probe the yet unexplored saturation regime at small-$x$ \cite{Gribov:1983ivg}. 
A significant part of these contemporary efforts, to understand the multi-dimensional structure of proton, revolves around the study of non-perturbative gluon-gluon correlators, especially, at small-$x$. 
Technically, the correlator is the bi-local off-forward hadronic matrix element of color field strength tensors at two different space points at some light-cone time.  
Parametrization of the correlator would give rise to the Generalized Transverse Momentum Dependent distributions (GTMDs). 
The GTMDs are functions of $x$, gluon transverse momentum $k_\perp$, transverse momentum transfer $\Delta_\perp$, and longitudinal momentum transfer aka skewness parameter $\xi$. 
Different projections, through the various independent combinations of  $k_\perp$,  $\Delta_\perp$, and spin vector, if there, $S$,  project out different GTMDs from the gluon-gluon correlator. 
All the GTMDs as well as their descendants $e.g.$  Transverse Momentum Dependent distributions (TMDs), Generalized Parton Distributions (GPDs), and Parton Distribution Functions (PDFs) are themselves non-perturbative objects and can be extracted only from the experiments. %
However their evolutions, both across the high scale $Q^2$ or small Bjorken-$x$, can be studied using the first principle perturbative Quantum Chromodynamics (QCD) setup. 

\vspace{0.5cm}

 While all the GTMDs can be thought of as the different scalar pieces of the non-perturbative color field strength correlator, they, however, evolve quite differently both along high scale $Q^2$ or small Bjorken variable $x$. 
Nucleon helicity non-flip distributions, $e.g.$, $H$-type GPDs follow the Balitsky-Fadin-Kuraev-Lipatov (BFKL) equation that stems from the single $\alpha_s \ln 1/x$ resummations \cite{Kuraev:1977fs,Balitsky:1978ic}.  
Considerable theoretical work has been done around this distribution that does not flip the helicity. 
However, the helicity-flip or spin-flip gluon $E_g$ GPDs or associated GTMDs are among the least explored, yet phenomenologically important, distributions. 
As GPD $E_g$ are associated with the nucleon helicity flip processes - it is a general belief that they are suppressed in the high energy. 
However, recently Hatta and Zhou have shown that GPD $E_g$ at vanishing skewness exhibits Regge behavior similar to the BFKL pomeron with identical intercept \cite{Hatta:2022bxn}.  
This is done by deriving the small-$x$ evolution equations for the two $F$-type spin-flip gluon GTMDs, the $f_{1,2}$ and $f_{1,3}$. 
In this article, we have analytically solved the equations and 
find small-$x$ asymptotics of the GTMDs $f_{1,2}$ (and $f_{1,3}$) that are related to the novel helicity flip processes.

\vspace{0.5cm} 
%

{\color{NavyBlue} {\it F-Type Spin-flip GTMDs}:~ }
\noindent
The gluon GTMDs can be defined through the parametrization of the off-forward bilocal correlator of the two gluon field strength tensors, 
\begin{eqnarray}
W^{[i,j]}_{\lambda, \lambda'}
&=& \left. \int \frac{d^2z_\perp dz^-}{\left(2\pi\right)^3P^+}~e^{ixP^+z^--ik_\perp.z_\perp}~\langle p', \lambda' | F_{a}^{+i}\left(-\frac{z}{2}\right)  {\cal U}^{[+]}_{\frac{z}{2},-\frac{z}{2}} F_{a}^{+j}\left(+\frac{z}{2}\right) {\cal U}^{[-]}_{-\frac{z}{2},\frac{z}{2}} |p, \lambda \rangle \right|_{z^+=0},  \label{GG}
\end{eqnarray}
where average proton momentum $P=(p+p')/2$ and momentum transfer $\Delta = p'-p$. While the transverse momentum transfer $\Delta_\perp$ is explicit in the expression, the longitudinal momentum transfer to the nucleon is presented through the skewness parameter $\xi$ as $ \xi = - \Delta^{+}/P^{+}$. 
The two gauge links ensure the gauge invariance of the color correlator. The prescription, however, is not unique and depends on the actual process under consideration. The two most used staple gauge links are the past pointing and future pointing gauge links \cite{Banu:2021cla}. The dipole distribution, that we are considering here, contains both. \\

$\bullet$ {\it F-type gluon GTMDs}:~ Contraction of $W^{[i,j]}$ by symmetric $\delta^{ij}$ will project the four complex (or equivalently eight real) $F$-type unpolarized gluon GTMDs. The unpolarized gluon TMD $f_{1}^{g}$ and gluon Siver's function $f_{1T}^{\perp g}$ are the forward limit ($\Delta_\perp=0$)  of two of these GTMDs. While $f_{1}^{g}$ is the distribution of unpolarized gluons in an unpolarized proton, the Siver function gives distributions of unpolarized gluons in a transversely polarized proton.  \\ 

$\bullet$ {\it G-type gluon GTMDs}:~ The anti-symmetric $i\epsilon^{ij}$ will project the $G$-type gluon GTMDs which are related to the distribution of circularly polarized gluons.  The gluon helicity TMD $g_{1L}^{g}$ for longitudinally polarized proton and worm-gear gluon TMD $g_{1T}^{g}$ for transversely polarized proton, both are the descendants of $G$-type gluon GTMDs. \\
 
$\bullet$ {\it H-type gluon GTMDs}:~ Projection by transverse spin  $S_\perp$ will single out the $H$-type gluon GTMDs. In the $\Delta_\perp=0$ limit, one recovers Boer-Mulder function $h_{1L}^{g}$ (distribution of linearly polarized gluon in an unpolarised proton) and others TMDs $e.g.$ $h_{1L}^{\perp g}$, $h_{1T}^{g}$, $h_{1T}^{\perp g}$ for linearly polarized gluons.  \\


\noindent 
In this article, we consider the $F$-type gluon GTMDs. 
%
%
%
\noindent
In the off-forward limit, at the leading twist,  $\delta^{ij}W^{[i,j]}$ are parameterized through the $F$-type GTMDs as below \cite{Mulders:2000sh,Meissner:2009ww,Lorce:2013pza,Boer:2016xqr,Boer:2018vdi,Boussarie:2019vmk,Bhattacharya:2018lgm}, 
\begin{eqnarray}
\delta^{ij}W^{[i,j]}_{\lambda, \lambda'}
&=& \frac{1}{2M}{\bar u}\left(p', \lambda'\right)\left[ F_{1,1} + i\frac{\sigma^{j+} k_\perp^j}{P^+}F_{1,2} + i\frac{\sigma^{j+}\Delta_{\perp}^{j}}{P^+}F_{1,3} + i\frac{\sigma^{ij}k_\perp^i \Delta_{\perp}^j}{M^2}F_{1,4}\right] u(p, \lambda).     \label{GG101}
\end{eqnarray}
All GTMDs, in the above expression, are functions of $(x,~k_\perp^2,~ \Delta_\perp^2,~  k_{\perp}\cdot\Delta_{\perp},~ \xi)$ and are in general complex functions. In the eikonal limit,  $\xi \ll 1$, one may write \cite{Boussarie:2019vmk}, for $n=1,3,4$, 
\begin{eqnarray}
F_{1,n} = f_{1,n} + i \frac{k_\perp \cdot \Delta_\perp}{M^2} \Tilde{f}_{1,n}, 
\end{eqnarray}
whereas,  for $n=2$, 
\begin{eqnarray}
F_{1,2} = \frac{k_\perp.\Delta_\perp}{M^2} f_{1,2} +  i \Tilde{f}_{1,2}.    \label{f12abc}
\end{eqnarray}
Clearly, in the off-forward limit ($\Delta_\perp \neq 0$) there are four complex $F$-type gluon GTMDs ($F_{1,n}$) or equivalently eight real GTMDs ($f_{1,n}$ and ${\tilde f}_{1,n}$). 
Integration over $k_\perp$ in Eq.\eqref{GG101}, would show up the two GPDs, spin non-flip $H_g$ and spin-flip $E_g$,
as follows, 
\begin{eqnarray}
\int d^2k_\perp W^{[i,i]}_{\lambda, \lambda'} = \frac{1}{2P^+} {\bar u}(p',\lambda') \left(H_g \gamma^+ + iE_g \frac{\sigma^{+\nu}\Delta_{\nu}}{2M}\right)u(p, \lambda).
\end{eqnarray}
Both GPDs are functions of $x,\Delta_\perp^2$ and  $\xi$. The first term is proportional to $\delta_{\lambda, \lambda'}$, while the second term is proportional to $\delta_{\lambda, -\lambda'}$ making the $H_g$ being the spin non-flip and $E_g$ being spin-flip distributions. 
%
%
The real GTMDs, $f_{1,1}$, $f_{1,2}$ and $f_{1,3}$  are related to the two GPDs by the following two integrals: 
\begin{eqnarray}
x H_g &=& \int d^2k_\perp ~ f_{1,1}(k_\perp), \label{Hggg} \\
x E_g &=& \int d^2k_\perp ~\left( - f_{1,1}(k_\perp) + \frac{k_\perp^2}{M^2} f_{1,2}(k_\perp)  + 2 f_{1,3}(k_\perp)\right).  \label{Eggg}
\end{eqnarray}
%
%
%
It's well known that in the dilute regime, where the saturation phenomena did not yet kick in, the unpolarised spin independent $f_{1,1}$ follow BFKL evolution, leading to 
\begin{eqnarray}
f_{1,1} \sim x G(x) \sim  \left(\frac{1}{x}\right)^{{\bar \alpha_s} 4\ln 2}~. 
\end{eqnarray}
Deep inside the saturation region, the distribution is naturally expected to follow the Balitsky-Kovchegov (BK) equation \cite{Balitsky:1995ub,Kovchegov:1999yj}. The dipole gluon TMD, so as the function $f_{1,1}$, at small-$k_\perp$ and at asymptotically small-$x$ is found to be proportional to $\ln(k_{\perp}^2/Q_s^2(x))$ where $Q_s(x)$ is the saturation scale \cite{Siddiqah:2018qey,Abir:2017mks}.  
Other than $f_{1,1}$, only ${\tilde f}_{1,2} = {\rm Im} \left(F_{1,2}^g\right)$ survives in the forward limit $\Delta_\perp = 0$. While the $f$-distributions follow pomeron evolution, the ${\tilde f}$-distributions are connected with odderons, $e.g.$ $\tilde{f}_{1,2}$ is known as spin-dependent odderons. Its $k_\perp$ moments are related to the three gluon correlators relevant for transverse single spin asymmetry \cite{Zhou:2013gsa}.  
In fact, for transversely polarized proton $\tilde{f}_{1,2}$ can be identified as gluon Sivers' function as, $x f_{1T}^{\perp, g}(x, k_\perp^2) \sim - 2 {\rm Im} \left(F_{1,2}^g\right)$. 
Odderons too satisfy the BFKL-like equation in the dilute regime with identical eigenfunctions and eigenvalues. However, as the $C$-odd initial conditions allow only the odd harmonics, the odderon intercept is found to be at zero \cite{Bartels:1999yt, Kovchegov:2003dm}. One may then expect that, 
\begin{eqnarray}
\tilde{f}_{1,2} \sim  \left(\frac{1}{x}\right)^{0}~. 
\end{eqnarray}
Access prospects of the gluon Sivers' function in the upcoming Electron-Ion Collider and its small-$x$ evolution has been studied recently \cite{Zheng:2018ssm, Yao:2018vcg}.  The function $F_{1,4}$ is associated with gluon orbital angular momentum and vanishes in the eikonal limit due to PT symmetry \cite{Bhattacharya:2022vvo,Bhattacharya:2023yvo}.

\vspace{0.5cm}

 {\color{NavyBlue}{\it Evolution of $f_{1,2}$}:~}
The small-$x$ evolution equation for ${f}_{1,2}$ as derived by Hatta and Zhou in \cite{Hatta:2022bxn}, is a non-linear integro-differential equation. In the dilute regime, where the non-linear term can be dropped, after some rearrangement of terms, the equation can be written as,  
\begin{eqnarray}
\frac{\partial }{\partial \ln(1/x)} \mathcal{F}_{1,2} \left(x,{ k}_\perp\right)=\frac{\bar{\alpha}_s}{\pi} \int \frac{d^2{k}'_{\perp}}{\left({k}_{\perp}-{k}'_{\perp}\right)^2} \left \{\mathcal{F}_{1,2}\left(x, {k}'_{\perp}\right) -\frac{k^2_\perp}{2{k'}^2_{\perp}} ~\mathcal{F}_{1,2}\left(x, {k}_{\perp}\right)+\frac{2\left({k}_{\perp}.{k}'_{\perp}\right)^2- {k}^2_{\perp}{k}'^2_{\perp}-k^4_\perp}{k_\perp^4}~\mathcal{F}_{1,2}\left(x,{k}'_\perp\right)  \right \}.  
\label{f12-B} 
\end{eqnarray}
The function ${\mathcal F}_{1,2}$ is defined for convenience and is related to $f_{1,2}$ as, 
\begin{eqnarray}
f_{1,2} = k_\perp^2 \frac{\partial^2}{\partial k_\perp^{i} \partial k_{\perp}^{i}} {\mathcal F}_{1,2}.  
\end{eqnarray}
We note that Eq.\eqref{f12-B} has IR poles at $k_\perp = 0$, $k'_\perp = 0$, and $k'_\perp = k_\perp $. The first two terms, on the right-hand side of Eq.\eqref{f12-B}, essentially constitute the BFKL kernel. 
%
%
As $k_{\perp}$ is a vector in the transverse plane, therefore $\mathcal{F}_{1,2}\left(x, {k}_{\perp}\right)$ in principle should be a function of the azimuthal angle $\phi_{k\Delta}$ between $k_{\perp}$ and $\Delta_{\perp}$.
%
Now one may study the eigenvalues of the integral operator as in Eq.\eqref{f12-B}, following the procedure outlined by Del Duca \cite{DelDuca:1995hf}. To begin with, we assume that the solution of the Eq.\eqref{f12-B} admits a Fourier series in the azimuthal angle:
\begin{eqnarray}
    \mathcal{F}_{1,2} \left(x, |k_\perp|, \phi_k \right) &=& \sum_{n=-\infty}^{\infty}~\mathcal{F}^{(n)}_{1,2}(x,k_{\perp}^2)~e^{in\phi_{k\Delta}}\label{f12-b1}, 
\end{eqnarray}
\noindent where $\mathcal{F}^{(n)}_{1,2}$~'s are the Fourier coefficients that depends on $x$ and $|k_{\perp}|$ (or on $k_\perp^2$) but not on $\phi_{k\Delta}$.
\noindent The inverse Mellin transformation for the variable $x$ gives the series solution as the integral over the complex 
variable $\gamma$ along a contour which is a straight vertical line in the complex plane,
\begin{eqnarray}
  \mathcal{F}^{(n)}_{1,2}(x,k_{\perp}^2) &=&  \int \frac{d\gamma}{2\pi i	} ~\left(\frac{1}{x}\right)^{\bar{\alpha}_s\chi_{1,2}(n,\gamma)}~\frac{k_{\perp}^{2\gamma}}{k_{\perp}^2},  \label{inter}
\end{eqnarray}
\noindent where one assume that the Mellin transform function of $\mathcal{F}^{(n)}_{1,2}$ consists of powers of transverse momentum, $k_{\perp}$ $i.e.$ it's a power law function of $k_\perp^2$. This essentially stems from the conformal structure of the kernel of evolution equation in Eq.\eqref{f12-B}. The power law structure makes the eigenfunction scale invariant. 

\vspace{0.5cm}

\noindent One may now move on to evaluate the eigenvalue $\chi_{1,2}(n,\gamma)$, corresponding to the above eigenfunction,
\begin{eqnarray}
\chi_{1,2}(n,\gamma) &=&\frac{1}{\pi} \int \frac{d^2{k}'_{\perp}}{\left({k}_{\perp}-{k}'_{\perp}\right)^2} \left \{\left(\frac{k'^2_\perp}{k^2_\perp}\right)^{\left(\gamma-1\right)} ~ e^{in\left(\phi_{k' \Delta}-\phi_{k \Delta} \right)} -\frac{k^2_\perp}{2{k}'^2_{\perp}} \right \}\nonumber \\
&& ~~~~~~~~~~~~~~~~~~ +\frac{1}{\pi} \int \frac{d^2{k}'_{\perp}}{\left({k}_{\perp}-{k}'_{\perp}\right)^2}\left(\frac{2\left({k}_{\perp}.{k}'_{\perp}\right)^2- {k}^2_{\perp}{k}'^2_{\perp}-k^4_\perp}{k_\perp^4} \right)~\left(\frac{k'^2_\perp}{k^2_\perp}\right)^{\left(\gamma-1\right)} e^{in\left(\phi_{k'\Delta}-\phi_{k\Delta} \right)}. \label{f12-c} 
\end{eqnarray}
\noindent
The first term in the above equation is the eigenvalue for the BFKL kernel $\chi_{BFKL}(n,\gamma)$.
%
%
After solving Eq.\eqref{f12-c}
the full eigenvalue $\chi_{1,2} (n, \gamma)$ is found to be, 
\begin{eqnarray}
\chi_{1,2} (n, \gamma) &=&  2 \psi(1) - \frac{1}{2}\psi \left(\gamma+\frac{|n|}{2}\right)
- \frac{1}{2}\psi \left(\gamma+\frac{|n|}{2}+2\right) - \frac{1}{2}\psi \left(-\gamma+\frac{|n|}{2}-1\right) -\frac{1}{2}\psi\left(-\gamma+\frac{|n|}{2}+1\right).  \label{A11}
\end{eqnarray}
\noindent While evaluating $\chi_{1,2}(n,\gamma)$  all IR divergences are mutually canceled leading to IR finite, divergence-free $\chi_{1,2}(n,\gamma)$ as presented in Eq.\eqref{A11}. This also shows that the evolution equation as given in, Eq.\eqref{f12-B},  is IR finite. 
%
%
\noindent  It is interesting to note that, unlike the BFKL eigenvalue, for which the saddle point is located at $Re\left(\gamma\right)=1/2$, 
the saddle point of $\chi_{1,2}(n,\gamma)$ is at $Re\left(\gamma\right)=-1/2$ for all $n$.
%
%
\begin{figure}
\begin{subfigure}[b]{.49\linewidth}
\centering
\includegraphics[width=\linewidth]{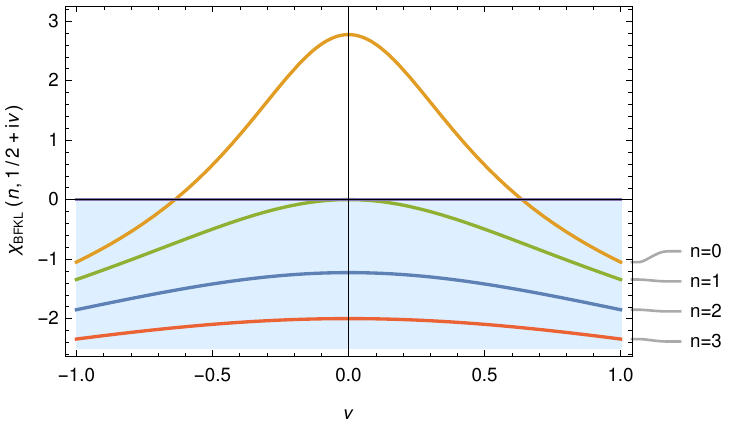}
\caption{$\chi_{BFKL}(n,\gamma)$ at the saddle point $\gamma = 1/2 + i0$.}\label{fig1a}
\end{subfigure}\hfill
\begin{subfigure}[b]{.49\linewidth}
\centering
\includegraphics[width=\linewidth]{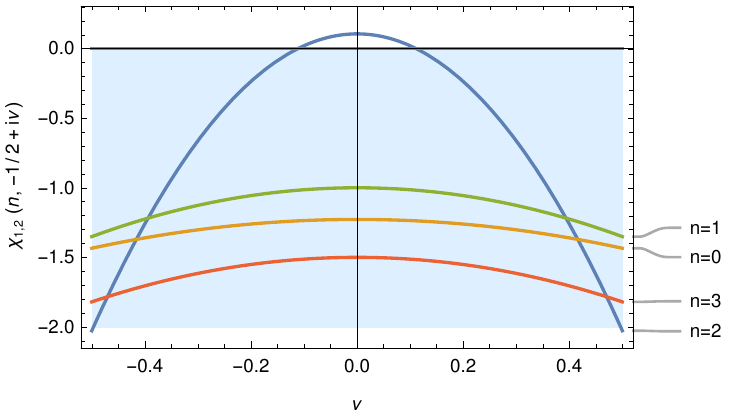}
\caption{$\chi_{1,2}(n,\gamma)$ at the saddle point $\gamma = -1/2 + i0$.}\label{fig1b}
\end{subfigure}%
\caption{Eigen values of the two kernels, at their respective saddle points, as a function of $\nu$.}
\end{figure}
\noindent
Taking $\gamma = -1/2 + i\nu$, 
\begin{eqnarray}
\chi_{1,2}\left(n,-\frac{1}{2}+i\nu\right)
&=& 2\psi\left(1\right)-Re\left[\psi\left(-\frac{1}{2}+\frac{|n|}{2}+i\nu\right)+\psi\left(\frac{3}{2}+\frac{|n|}{2}-i\nu\right)\right], 
\end{eqnarray}
we may now expand $\chi_{1,2}(n,\nu)$ around $\nu=0$, and evaluate the integral in Eq.\eqref{inter} for different values of $n$.

\vspace{0.5cm}

%

$\bullet$ {\it Special case $|n|=0$}:~
%
%
%
If the transverse momentum of gluons is not too large $i.e.$ $k_\perp \sim \Lambda$, which still is larger than $\Lambda_{QCD}$, one may evaluate the $\nu$-integral in Eq.\eqref{inter} in the diffusion approximation and get the amplitude for first or `radial' harmonic (corresponds to $n=0$). However, 
%
%
 the leading behavior of ${\cal F}_{1,2}^{(0)}$ is found to be $ \sim x^{4\left(1-\ln 2\right)\bar{\alpha}_s }$, and this term will not survive in the high energy.

\vspace{0.5cm}

$\bullet$ {\it Special case $|n|=1$ and all other odd harmonics}:~
%
%
%
As the function $f_{1,2}$ is even under the transformation $\Delta_\perp \rightarrow -\Delta_\perp$ or equivalently it depends on the azimuthal angle $\theta_{k\Delta}$ only through  $|k_\perp.\Delta_\perp|$, all the odd harmonics,  $|n|=1,3 ..$ will vanish identically from the very beginning.  
\vspace{0.5cm}

$\bullet$ {\it Special case $|n|=2$}:~ This is the first and only mode that survives in the high energy. After $\nu$ integration in the saddle point approximation one gets,  
%
%
\begin{eqnarray}
{\cal F}_{1,2}^{(2)}(x,k^2) =  \frac{1}{\pi }  \frac{\Lambda}{|\textit{k}_\perp|^3}\left(\frac{\pi}{14\left(\zeta(3)-\frac{16}{27}\right)~\bar{\alpha}_s \ln{(1/x)}}\right)^{1/2} {\left(\frac{1}{x}\right)}^{\bar{\alpha}_s\left(4\ln{2}-8/3\right)}\exp\left \{-\frac{\ln^2(\textit{k}_\perp/\Lambda)}{14\left(\zeta(3)-\frac{16}{27}\right)\bar{\alpha}_s\ln(1/x)}\right \} .  \nonumber \\
\end{eqnarray}
\noindent
Unlike $|n|=0$ and $|n|=1$, one may now observe that, 
\begin{eqnarray}
{\cal F}_{1,2}^{(2)}(x,k^2) \sim \left(\frac{1}{x}\right)^{(4\ln 2 - 8/3){\bar \alpha}_{s}}.
\end{eqnarray}
As $4\ln 2 - 8/3=0.106$ is positive, the term will survive in the small-$x$ and govern the leading small-$x$ behavior of   ${\cal F}_{1,2} (x, k_\perp)$. 
%
%
%
%
%
%
%
%
%
%
For all other values of $n$,  other than $n=2$,  the $\chi_{1,2}$~'s are finite and negative at $\nu=0$ (See Fig.\ref{fig1b}). In the small-$x$ limit, only the term corresponding to $n=2$ will survive, leading to, 
\begin{eqnarray}
\mathcal{F}_{1,2} \left(x, |k_\perp|, \phi_{k\Delta} \right) \sim \frac{\Lambda}{|k_{\perp}|^3} \left(\frac{1}{x}\right)^{\left(4\ln 2-8/3\right)\bar{\alpha}_s} {\cal S}_{2}(x,k_\perp^2)~~2\cos(2\phi_{k\Delta}), 
\end{eqnarray}
%
%
where, 
\begin{eqnarray}
   {\cal S}_{2}(x,k_\perp^2)=\left(\frac{\pi}{\tau_2\bar{\alpha}_s\ln(1/x)}\right)^{1/2}\exp \left\{-\frac{\ln^2(k_\perp^2/\Lambda^2)}{4\tau_2\bar{\alpha}_s\ln(1/x)}\right\},
\end{eqnarray}
with $4\ln2-8/3=0.106$ and $\tau_2$ being $14(\zeta(3)-16/27)$. Interestingly in the leading order, we see that the evolution of $\mathcal{F}_{1,2}$ has an azimuthal, $\phi_{k\Delta}$ dependence. Unlike BFKL, the surviving solution corresponds to conformal spin $n=2$. This leads to an explicit $2\cos 2\phi_{k\Delta}$ dependence in the GTMDs. 
As the function $f_{1,2}$ comes with an additional $\cos\phi_{k\Delta}$ from the prefactor $k_{\perp}.\Delta_{\perp}$ as shown in Eq.\eqref{f12abc} - this will lead to, 
\begin{eqnarray}
{\rm Re}(F_{1,2}) \sim \left(\frac{1}{x}\right)^{\alpha_s\left(4\ln2-8/3\right)} \left( \cos 3\phi_{k\Delta} + \cos \phi_{k\Delta} \right). \nonumber 
\end{eqnarray}
We note here that, the $\cos2\phi_{k\Delta}$ correlation between $k_\perp$ and $\Delta_\perp$ contained in $f_{1,1}$ - leading to `elliptic GTMDs' have been studied earlier \cite{Hatta:2016dxp}. 
%

\vspace{0.5cm}

{\color{NavyBlue} {\it Evolution of} $f_{1,3}$:~}
While the evolution equation for ${\cal F}_{1,2}$ as presented in Eq.\eqref{f12-B} is a closed equation, the evolution equation for ${\cal F}_{1,3}$ is not a closed one. The evolution of ${\cal F}_{1,3}$ not just depends on itself but depends on
${\cal F}_{1,2}$ as well, 
\begin{eqnarray}
\frac{\partial }{\partial Y} \mathcal{F}_{1,3} \left({ k}_\perp\right)=\frac{\bar{\alpha}_s}{\pi} \int \frac{d^2\it{k}'_{\perp}}{\left(\it{k}_{\perp}-\it{k}'_{\perp}\right)^2} \left 
\{ {\cal F}_{1,3}(k'_\perp) -\frac{k^2_\perp}{2 k'^2_{\perp}} ~\mathcal{F}_{1,3}\left({k}_\perp\right)   -
\frac{\left(\it{k}_{\perp}.\it{k}'_{\perp}\right)^2- {k}^2_{\perp}{k}'^2_{\perp}}{k_\perp^2M^2}~\mathcal{F}_{1,2}\left(k'_\perp\right)  \right \}. 
\label{f13-two}
\end{eqnarray}
We assume that $\mathcal{F}_{1,3}$ has the following form: 
\begin{eqnarray}
    {\mathcal{F}_{1,3}}(x,{ k}_{\perp})={\cal C}_1\phi_1(x,{ k}_{\perp})+{\cal C}_2\frac{k_{\perp}^2}{M^2}\phi_2(x,{ k}_{\perp}), \label{f13-twoA}
\end{eqnarray}
where both $\phi_1$ and $\phi_2$ are some regular functions of $x$ and $k_\perp$. The coefficients ${\cal C}_{1,2}$  are 
dimensionless constants. Substituting Eq.\eqref{f13-twoA} in Eq.\eqref{f13-two}, rearranging, and then equating the mass-independent and mass-dependent parts to be zero, would make $\phi_1$ and $\phi_2$ to be identified with ${\cal F}_{1,1}$ and ${\cal F}_{1,2}$ with ${\cal C}=1$ and ${\cal C}_2=-1/2$ (details are in the appendix). 
Therefore, 
\begin{eqnarray}
 {\mathcal{F}_{1,3}}(x,{ k}_{\perp})={\cal F}_{1,1}(x,{ k}_{\perp})-\frac{k_{\perp}^2}{2M^2}{\cal F}_{1,2}(x,{ k}_{\perp}). 
\end{eqnarray}
The small-$x$ asymptotics of the two functions on the right side of the above equation are now known, 
\begin{eqnarray}
{\cal F}_{1,1}(x,{ k}_{\perp}) \sim \left( \frac{1}{x} \right)^{4\ln 2 {\bar \alpha}_s};  ~~~~~~~~ 
{\cal F}_{1,2}(x,{ k}_{\perp}) \sim \left( \frac{1}{x} \right)^{(4\ln 2-8/3){\bar \alpha}_s}.  
\end{eqnarray}
%
%
%

\vspace{0.5cm}

{\color{NavyBlue} {\it Summary and Outlook:~}}
Until a few years ago, it was virtually unknown how to measure gluon GTMDs. 
Only recently it was shown that gluon GTMDs, can in principle be accessed via diffractive di-jet production in the deep-inelastic electron-proton and electron-ion collisions \cite{Hatta:2016dxp, Hatta:2016aoc, Ji:2016jgn}. At about the same time, it was proposed that the GTMDs, especially the `elliptic' GTMD can be accessed via virtual photon-nucleus quasi-elastic scattering \cite{Zhou:2016rnt} and also in proton-nucleus collisions \cite{Hagiwara:2017ofm}. The three GTMDs $\tilde{f}_{1,1}$, $\tilde{f}_{1,2}$ and $\tilde{f}_{1,3}$, which describe how odderons couple to generic spin-1/2 hadrons, have been studied in the off-forward kinematics, and found to be accessible via exclusive pion production in the deep-inelastic scatterings \cite{Boussarie:2019vmk}. These studies on accessing GTMDs, so far, cover either the unpolarized $f_{1,1}$, odderon inspired ${\tilde f}_{1,i}$ (for their connection to gluon Siver function \cite{Boussarie:2019vmk}) or $F_{1,4}$ due to its close connection to gluon orbital angular momentum \cite{Bhattacharya:2022vvo}. The two GTMDs, $f_{1,2}$ and $f_{1,3}$, that we studied here, have been relatively less explored in the phenomenological context.    \\ 

Both the positive intercept and the nontrivial angular correlation between $k_{\perp}$ and ${\Delta_\perp}$, the one we found in this study, is likely to have phenomenological consequences and relevant observables, within the kinematic and detector reach of the upcoming Electron-Ion collider. The proton recoil momentum $\Delta_\perp$ can be directly measured at the EIC, thanks to the planned installation of Roman pots and the off-momentum detector to be placed very close to the beamline to track the recoil proton. \\

GTMD $f_{12}$ appears in some results in the literature (e.g. \cite{Hagiwara:2020mqb}), however finding a process in which $f_{12}$ plays the dominant role is a
challenge - and open to the community to explore as of now.  \\

Many efforts, especially in the theory front, have been made in the last few years to explore phenomenology aimed at nucleon tomography in terms of GTMDs for current and future experiments. 
%
%
%
%
%
%
While the TMDs and GTMDs studies - especially their small-$x$ evolutions in the quarks sector have been studied a lot in recent times  \cite{Kovchegov:2021iyc,Santiago:2023rfl}, the gluons PDFs, especially, the helicity PDFs (hPDFs) \cite{Adamiak:2023yhz}, gluon  TMDs \cite{Boer:2015pni,Chakrabarti:2023djs}, related non-perturbative parameters $e.g.$ jet quenching parameter \cite{Abir:2015qva} or gluon GTMDs \cite{Bhattacharya:2023yvo} are relatively less explored objects. Fresh new approach to the nucleon tomography $e.g.$ based on 
nucleon energy correlators \cite{Liu:2022wop,Liu:2023aqb} or correlations of di-hadron productions between the current fragmentation region (CFR) and target fragmentation region (TFR) in DIS \cite{Guo:2023uis} are now start appearing.  \\

In this work, we have analytically solved the small-$x$ evolution equation for spin-flip gluon GTMD $f_{1,2}$ and $f_{1,3}$. Key results are as follows: 

\vspace{0.3cm}

$(a)$  The evolution equation for ${\rm Re}(F_{1,2})$ carries IR singular terms. We have shown that all IR divergences, from different terms, mutually cancel making the equation a self-consistent and closed equation. The only known examples of such IR-safe equations, within small-$x$ physics, are the celebrated BFKL equation (or BK equation) and the Odderon equation.     \\

$(b)$ The intercept for ${\rm Re}(F_{1,2})$ is found to be positive, 
\begin{eqnarray}
{\rm Re}(F_{1,2})  \sim \left( \frac{1}{x} \right)^{(4\ln 2-8/3){\bar \alpha}_s},  \nonumber  
\end{eqnarray}
which implies that it is expected to dominate over the gluon Siver function in the small-$x$ limit. 
This may directly impact the modeling of unpolarised GTMDs and associated spin-flip processes. \\

$(c)$ Unlike, BFKL or Odderon evolution equations, the surviving solution corresponds to conformal spin $n=2$. This leads to an explicit $\cos 3\phi_{k\Delta} + \cos \phi_{k\Delta}$ azimuthal dependence in the GTMDs - may translate to angular correlation observable for slip-flip processes.   \\ 

$(d)$ There are two broad classes of GTMDs, the first set that survives in the forward limit, and the second set that does not survive in the forward limit. The second class of GTMDs is realized only in the off-forward limit $\Delta_\perp \neq 0$. The present paper is the first result on the small-$x$ asymptotics of any gluon GTMDs that belong to the second class. \\

\vspace{0.5cm}


{\color{NavyBlue} {\it Acknowledgments}:~ } 
We thank Yoshitaka Hatta for the valuable discussions and suggestions throughout this work. The work is supported by the DST - Govt. of India through SERB-MATRICS Project Grant No. MTR/2019/001551.


\section{ Appendix }
%

\subsection{{\color{NavyBlue} Derivation of eigenvalue $\chi_{1,2}$}}
\noindent The eigenvalue $\chi_{1,2}$ defined as, 
\begin{eqnarray}
\frac{1}{\pi} \int \frac{d^2\textit{k}'_{\perp}}{\left(\textit{k}_{\perp}-\textit{k}'_{\perp}\right)^2} \left[ \left( \frac{2\left(\textit{k}_{\perp}.\textit{k}'_{\perp}\right)^2-\textit{k}^2_{\perp} {k}'^2_{\perp}}{k_\perp^4}\right)~ {k}'^{2(\gamma-1)}_{\perp}e^{in\phi_{k'\Delta}} -\frac{k^2_\perp}{2\textit{k}'^2_{\perp}} ~{k}^{2(\gamma-1)}_{\perp}e^{in\phi_{k\Delta}} \right]   = \chi_{1,2} 
\left(n, \gamma \right) ~  {k}^{2(\gamma-1)}_{\perp}e^{in\phi_{k\Delta}} \label{A0}.
\end{eqnarray}
\noindent Below we present an outline to evaluate the eigenvalue $\chi_{1,2}(n,\gamma)$, 
\begin{eqnarray}
\chi_{1,2}(n,\gamma) &=&\frac{1}{\pi} \int \frac{d^2{k}'_{\perp}}{\left({k}_{\perp}-{k}'_{\perp}\right)^2} \left \{\left(\frac{k'^2_\perp}{k^2_\perp}\right)^{\left(\gamma-1\right)} ~ e^{in\left(\phi_{k' \Delta}-\phi_{k \Delta} \right)} -\frac{k^2_\perp}{2{k}'^2_{\perp}} \right \}\nonumber \\
&& ~~~~~~~~~~~~~~~~~~ +\frac{1}{\pi} \int \frac{d^2{k}'_{\perp}}{\left({k}_{\perp}-{k}'_{\perp}\right)^2}\left(\frac{2\left({k}_{\perp}.{k}'_{\perp}\right)^2- {k}^2_{\perp}{k}'^2_{\perp}-k^4_\perp}{k_\perp^4} \right)~\left(\frac{k'^2_\perp}{k^2_\perp}\right)^{\left(\gamma-1\right)} e^{in\left(\phi_{k'\Delta}-\phi_{k\Delta} \right)}, \label{f12-cd} \\ 
&\equiv & \chi_{BFKL}(n,\gamma) + \chi_{1,2}^{*}(n,\gamma).
\end{eqnarray}
\noindent
The first term in the above equation is the eigenvalue for the BFKL kernel.
\begin{eqnarray}
\chi_{{}_{BFKL}}(n,\gamma) = 2\psi(1) - \psi \left(\gamma+\frac{|n|}{2}\right) - \psi \left(1-\gamma+\frac{|n|}{2}\right).
\end{eqnarray}
We now solve the second term of the above equation as,
\begin{eqnarray}
\chi_{{}_{12}}^{*}(n,\gamma) &=&\frac{1}{\pi} \int \frac{d^2{k}'_{\perp}}{\left({k}_{\perp}-{k}'_{\perp}\right)^2}\left(\frac{2\left({k}_{\perp}.{k}'_{\perp}\right)^2- {k}^2_{\perp}{k}'^2_{\perp}-k^4_\perp}{k_\perp^4} \right)~\left(\frac{k'^2_\perp}{k^2_\perp}\right)^{\left(\gamma-1\right)} e^{in\left(\phi_{k'\Delta}-\phi_{k\Delta} \right)} 
\end{eqnarray}
We now define $t={k}'^2_{\perp}/{k}^2_{\perp}$ for convenience and rewrite the first term of Eq.\eqref{f12-cd} as, 
\begin{eqnarray}
\chi_{{}_{1,2}}^{*}(n,\gamma) &=& \frac{1}{2\pi}\int dt \int d\phi_{k'} \frac{2 t \cos^2(\phi_{k'\Delta}-\phi_{k\Delta})-t-1}{1+t-2\sqrt{t}\cos(\phi_{k'}-\phi_{k\Delta})}~e^{in(\phi_{k'\Delta}-\phi_{k\Delta})} t^{\gamma-1},
\end{eqnarray}
which can be further written, taking $z=\exp(i(\phi_{k'}-\phi_{k}))$, as, 
\begin{eqnarray}
\chi_{{}_{1,2}}^{*}(n,\gamma) =\frac{i}{2\pi}\int \frac{dt}{\sqrt{t}} \int dz \frac{t\left(z^2+1\right)^2/2-tz^2-z^2}{z^2\left(z-\sqrt{t}\right)\left(z-\frac{1}{\sqrt{t}}\right)}z^{|n|} t^{\gamma-1}.
\label{chi-hash-B}
\end{eqnarray}
The first term of Eq.\eqref{f12-cd} has IR singularity at $k_\perp' = k_\perp$ and also at $k_\perp = 0$. Both will be mapped in Eq.\eqref{chi-hash-B} as singularities at $t=1$ and at $t=\infty$ respectively. 
The $z$-integral runs clockwise along a unit circle around the origin in the complex $z$-plane. Also as the angular integration is an even function of $n$, therefore, can be written only as a function of $|n|$. We now perform the $z$-integral by methods of residue for poles at $z=\sqrt{t}$ and also at $z=1/\sqrt{t}$. We note here that, for $n=0$ and $n=1$, there is one more pole at $z=0$. Contributions from the $z=0$ pole are, however, found to be zero for all values of $\gamma$. After performing $z$-integration and $t$-integration, one finally arrive at,  
%
%
%
%
%
%
%
%
\begin{eqnarray}
\chi_{{}_{12}}^{*}(n,\gamma) = \left[\frac{\gamma}{(n/2)^2-\gamma^2}+\frac{\gamma+1}{(n/2)^2 -(\gamma+1)^2} \right].
\end{eqnarray}
\noindent
Complete eigen value can now be written by adding $\chi_{{}_{BFKL}}(n,\gamma)$ and $\chi_{{}_{12}}^{*}(n,\gamma) $ together as, 
\begin{eqnarray}
\chi_{1,2} (n, \gamma) &=&  2 \psi(1) - \frac{1}{2}\psi \left(\gamma+\frac{|n|}{2}\right)
- \frac{1}{2}\psi \left(\gamma+\frac{|n|}{2}+2\right) - \frac{1}{2}\psi \left(-\gamma+\frac{|n|}{2}-1\right) -\frac{1}{2}\psi\left(-\gamma+\frac{|n|}{2}+1\right).  \label{A11}
\end{eqnarray}
\noindent While calculating $\chi_{1,2}(n,\gamma)$  all IR divergences are mutually canceled leading to IR finite, divergence-free $\chi_{1,2}(n,\gamma)$. This also shows that the equation under consideration is IR finite. 

\subsection{{\color{NavyBlue} Saddle point at $\gamma = -1/2$}}

\noindent To identify the saddle point, one needs to find the maxima of $\chi_{1,2}$,
\begin{eqnarray}
\frac{d}{d\gamma}\chi_{1,2}(n,\gamma) = - \frac{1}{2}\psi^{(1)} \left(\gamma+\frac{|n|}{2}\right)
- \frac{1}{2}\psi^{(1)} \left(\gamma+\frac{|n|}{2}+2\right) + \frac{1}{2}\psi^{(1)} \left(-\gamma+\frac{|n|}{2}-1\right) +\frac{1}{2}\psi^{(1)}\left(-\gamma+\frac{|n|}{2}+1\right) = 0. \label{saddle}
\end{eqnarray}
\noindent  From Eq.\eqref{saddle} one may observe that the saddle point of $\chi_{1,2}(n,\gamma)$ is at $Re\left(\gamma\right)=-1/2$ for all $n$.
%
%
\noindent
Taking $\gamma = -1/2 + i\nu$, 
\begin{eqnarray}
\chi_{1,2}\left(n,-\frac{1}{2}+i\nu\right)
&=& 2\psi\left(1\right)-Re\left[\psi\left(-\frac{1}{2}+\frac{|n|}{2}+i\nu\right)+\psi\left(\frac{3}{2}+\frac{|n|}{2}-i\nu\right)\right],
\end{eqnarray}
\noindent we may now expand $\chi_{1,2}(n,\gamma)$ around $\nu=0$. Below, we have jotted down expansion of $\chi_{1,2}\left(n,-\frac{1}{2}+i\nu\right)$ for some initial values of $n$: 
\begin{center}
\begin{tabular}{|c | c |} 
 \hline
$|n|$ & $\chi_{1,2}\left(n,-\frac{1}{2} + i\nu \right)$  \\ [1ex] 
 \hline 
 0 & $- 4(1-\ln 2) - [14\zeta (3) - 16]\nu^2 $     \\  
 \hline
 1 & $-1 - \left[  2 \zeta (3) - 1\right] \nu^2   $  \\
 \hline
 2 & $(4 \ln 2 - 8/3) - 14 \left[\zeta(3)-16/27\right]\nu^2  $  \\
 \hline
 3 & $- 3/2  - \left[2\zeta(3)-9/8\right]\nu^2$ \\  [1ex] 
 \hline 
\end{tabular}
\end{center}
\noindent

\vspace{0.5cm}

\subsection{{\color{NavyBlue} Saddle point integration over $\nu$ for $n=0,1,2$}}
If the transverse momentum of gluons is not too large $i.e.$ $k_\perp \sim \Lambda$, which still is larger than $\Lambda_{QCD}$, one may evaluate the $\nu$-integral, in the diffusion approximation, and get the amplitude, 
\begin{eqnarray}
  \mathcal{F}^{(n)}_{1,2}(x,k_{\perp}^2) &=& \sum_{n=-\infty}^{\infty} \int \frac{d\gamma}{2\pi i	} ~\left(\frac{1}{x}\right)^{\bar{\alpha}_s\chi_{12}(n,\gamma)}~\frac{k_{\perp}^{2\gamma}}{k_{\perp}^2},  \label{inter}
\end{eqnarray}

$\bullet$ {\it Special case $|n|=0$}:~
\begin{eqnarray}
\chi_{1,2}\left(0,-\frac{1}{2} + i\nu\right) = - 4(1-\ln 2) - [14\zeta (3) - 16]\nu^2,
\end{eqnarray}
The first or `radial' harmonic (corresponds to $n=0$) amplitude is found to be, 
\begin{eqnarray}
{\cal F}_{1,2}^{(0)}(x,k^2) =    \frac{\Lambda}{|\textit{k}_\perp|^3}\left(\frac{\pi}{\left(14\zeta(3)- 16\right)~\bar{\alpha}_s \ln{(1/x)}}\right)^{1/2} {\left(\frac{1}{x}\right)}^{-4\left(1-\ln 2\right)\bar{\alpha}_s } \exp\left \{-\frac{\ln^2(\textit{k}_\perp/\Lambda)}{\left(14\zeta(3)-16\right)\bar{\alpha}_s\ln(1/x)}\right \} .  
\end{eqnarray}
As the leading behavior of ${\cal F}_{1,2}^{(0)}$ is $ \sim x^{4\left(1-\ln 2\right)\bar{\alpha}_s }$, this term will also not survive in the high energy.

\vspace{0.5cm}

$\bullet$ {\it Special case $|n|=1$ and all other odd harmonics}:~
%
%
For $|n|=1$, the expansion of $\chi_{1,2}$ is, 
\begin{eqnarray}
\chi_{1,2}\left(1,-\frac{1}{2} + i\nu\right) = -1 - \left[ 2\zeta(3) - 1\right] \nu^2.
\end{eqnarray}
This will lead to,  
\begin{eqnarray}
    {\cal F}_{1,2}^{(1)}(x,k^2) =    \frac{\Lambda}{|\textit{k}_\perp|^3}\left(\frac{\pi}{\left(2\zeta(3)- 1\right)~\bar{\alpha}_s \ln{(1/x)}}\right)^{1/2} {\left(\frac{1}{x}\right)}^{-\bar{\alpha}_s } \exp\left \{-\frac{\ln^2(\textit{k}_\perp/\Lambda)}{\left(2\zeta(3)-1\right)\bar{\alpha}_s\ln(1/x)}\right \}. 
\end{eqnarray}
Again, the leading small-$x$ behavior of ${\cal F}_{1,2}^{(1)}$ is $ \sim x^{\bar{\alpha}_s }$, therefore this term will not survive in the high energy. In fact, as the function $f_{1,2}$ is even under the transformation $\Delta_\perp \rightarrow -\Delta_\perp$ or equivalently it depends on the azimuthal angle $\theta_{k\Delta}$ only through  $|k_\perp.\Delta_\perp|$, all the odd harmonics,  $|n|=1,3 ..$ will vanish identically from the very beginning.  
\vspace{0.5cm}

$\bullet$ {\it Special case $|n|=2$}:~
For $n=2$ the expansion around $\nu=0$ is, 
\begin{eqnarray}
\chi_{1,2}\left(2,-\frac{1}{2} + i\nu\right) = \left(4 \ln 2-\frac{8}{3}\right) - 14\left[\zeta (3) - \frac{16}{27} \right]\nu^2. 
\end{eqnarray}
This will lead to the following amplitude for ${\cal F}_{1,2}^{(2)}(x,k^2)$, 
\begin{eqnarray}
{\cal F}_{1,2}^{(2)}(x,k^2) =  \frac{1}{\pi }  \frac{\Lambda}{|\textit{k}_\perp|^3}\left(\frac{\pi}{14\left(\zeta(3)-\frac{16}{27}\right)~\bar{\alpha}_s \ln{(1/x)}}\right)^{1/2} {\left(\frac{1}{x}\right)}^{\bar{\alpha}_s\left(4\ln{2}-8/3\right)}\exp\left \{-\frac{\ln^2(\textit{k}_\perp/\Lambda)}{14\left(\zeta(3)-\frac{16}{27}\right)\bar{\alpha}_s\ln(1/x)}\right \} .  \nonumber \\
\end{eqnarray}
\noindent
Unlike $|n|=0$ and $|n|=1$, one may observe that, 
\begin{eqnarray}
{\cal F}_{1,2}^{(2)}(x,k^2) \sim \left(\frac{1}{x}\right)^{(4\ln 2 - 8/3){\bar \alpha}_{s}}.
\end{eqnarray}
As $4\ln 2 - 8/3=0.106$ is positive, the term will survive in the small-$x$ and govern the leading small-$x$ behavior of   ${\cal F}_{1,2} (x, k_\perp)$. 

\vspace{0.5cm}

\subsection{{\color{NavyBlue} {\it Evolution of} $f_{1,3}$~}}

%
While the evolution equation for ${\cal F}_{1,2}$ as presented is a closed equation, the evolution equation for ${\cal F}_{1,3}$ is not a closed one. The evolution of ${\cal F}_{1,3}$ not just depends on itself but depends on
${\cal F}_{1,2}$ as well. 
\begin{eqnarray}
\frac{\partial }{\partial Y} \mathcal{F}_{1,3} \left({ k}_\perp\right)=\frac{\bar{\alpha}_s}{\pi} \int \frac{d^2\it{k}'_{\perp}}{\left(\it{k}_{\perp}-\it{k}'_{\perp}\right)^2} \left 
\{ {\cal F}_{1,3}(k'_\perp) -\frac{k^2_\perp}{2 k'^2_{\perp}} ~\mathcal{F}_{1,3}\left({k}_\perp\right)   -
\frac{\left(\it{k}_{\perp}.\it{k}'_{\perp}\right)^2- {k}^2_{\perp}{k}'^2_{\perp}}{k_\perp^2M^2}~\mathcal{F}_{1,2}\left(k'_\perp\right)  \right \}. 
\label{f13-two}
\end{eqnarray}
We assume that $\mathcal{F}_{1,3}$ has the following form: 
\begin{eqnarray}
    {\mathcal{F}_{1,3}}(x,{ k}_{\perp})={\cal C}_1\phi_1(x,{ k}_{\perp})+{\cal C}_2\frac{k_{\perp}^2}{M^2}\phi_2(x,{ k}_{\perp}).\label{f13-twoA}
\end{eqnarray}
where both $\phi_1$ and $\phi_2$ are some regular functions of $x$ and $k_\perp$. The coefficients ${\cal C}_{1,2}$  are 
dimensionless constants. 
Substituting Eq.\eqref{f13-twoA} in Eq.\eqref{f13-two} and rearranging,
\begin{eqnarray}
    &&{\cal C}_{1}\frac{\partial }{\partial Y}~\phi_1(x,{k}_{\perp})- {\cal C}_{1}\frac{\bar{\alpha}_s}{\pi} \int \frac{d^2\textit{k}'_{\perp}}{\left(\textit{k}_{\perp}-\textit{k}'_{\perp}\right)^2}~\left[\phi_1(x,{k}'_{\perp}) -\frac{{{k^2_{\perp}}}}{2{k'^2_{\perp}}}~\phi_1(x,{ k}_{\perp})\right] \nonumber\\
    &&  +\frac{k_{\perp}^2}{M^2}\left[{\cal C}_{2}\frac{\partial }{\partial Y} \phi_2(x,{k}_{\perp}) 
   -\frac{\bar{\alpha}_s}{\pi} \int \frac{d^2\textit{k}'_{\perp}}{\left(\textit{k}_{\perp}-\textit{k}'_{\perp}\right)^2}~\left\{{\cal C}_{2}\frac{k'^2_\perp}{k^2_\perp}~\phi_2(x,{ k}'_{\perp})-\frac{\left(\textit{k}_{\perp}.\textit{k}'_{\perp}\right)^2- {k}^2_{\perp}{k}'^2_{\perp}}{(k_\perp^2)^2}\mathcal{F}_{1,2}\left(x,k'_{\perp}\right)- {\cal C}_{2}\frac{{{k^2_{\perp}}}}{2k'^2_{\perp}}~\phi_2(x,{ k}_{\perp})\right\}\right]=0.\nonumber\\
\end{eqnarray}
In the above equation, equating the coefficients of the mass-independent part to zero, we can write
\begin{eqnarray}
    \frac{\partial }{\partial Y}~\phi_1(x,{k}_{\perp})- \frac{\bar{\alpha}_s}{\pi} \int \frac{d^2\textit{k}'_{\perp}}{\left(\textit{k}_{\perp}-\textit{k}'_{\perp}\right)^2}~\left[\phi_1(x,{k}'_{\perp}) -\frac{{{k^2_{\perp}}}}{2{k'^2_{\perp}}}~\phi_1(x,{ k}_{\perp})\right]=0.
\end{eqnarray}
Clearly the evolution of $\phi_1$ is BFKL type and thus, we identify $\phi_1$ to be ${\cal F}_{1,1}$ with ${\cal C}_{1}=1$. 
Now, equating the coefficient of the mass-dependent part to zero,
\begin{eqnarray}
    {\cal C}_{2}\frac{\partial }{\partial Y} \phi_2(x,{k}_{\perp}) 
   -\frac{\bar{\alpha}_s}{\pi} \int \frac{d^2\textit{k}'_{\perp}}{\left(\textit{k}_{\perp}-\textit{k}'_{\perp}\right)^2}~\left\{{\cal C}_{2}\frac{k'^2_\perp}{k^2_\perp}~\phi_2(x,{ k}'_{\perp})-\frac{\left(\textit{k}_{\perp}.\textit{k}'_{\perp}\right)^2- {k}^2_{\perp}{k}'^2_{\perp}}{(k_\perp^2)^2}\mathcal{F}_{1,2}\left(x,k'_{\perp}\right)-{\cal C}_{2}\frac{{{k^2_{\perp}}}}{2k'^2_{\perp}}~\phi_2(x,{ k}_{\perp})\right\}=0. \label{A!}
\end{eqnarray}
\noindent
One may observe that $\phi_2={\cal F}_{1,2}$ and ${\cal C}_{2}=-1/2$ would satisfy the above equation since,
\begin{eqnarray}
     \frac{\partial }{\partial Y} {\cal F}_{1,2}(x,{k}_{\perp}) 
   -\frac{\bar{\alpha}_s}{\pi} \int \frac{d^2\textit{k}'_{\perp}}{\left(\textit{k}_{\perp}-\textit{k}'_{\perp}\right)^2}~\left\{\frac{2\left(\textit{k}_{\perp}.\textit{k}'_{\perp}\right)^2- {k}^2_{\perp}{k}'^2_{\perp}}{(k_\perp^2)^2}\mathcal{F}_{1,2}\left(x,k'_{\perp}\right)-\frac{{{k^2_{\perp}}}}{2k'^2_{\perp}}~{\cal F}_{1,2}(x,{ k}_{\perp})\right\}=0.
\end{eqnarray}
\noindent
Therefore, 
\begin{eqnarray}
 {\mathcal{F}_{1,3}}(x,{ k}_{\perp})={\cal F}_{1,1}(x,{ k}_{\perp})-\frac{k_{\perp}^2}{2M^2}{\cal F}_{1,2}(x,{ k}_{\perp}).
\end{eqnarray}

\end{document}